\documentclass{ws-ijmpe}
\usepackage[super]{cite}
\usepackage{xcolor}
\usepackage{amssymb}
\usepackage[verbose,hypertexnames=false]{hyperref}
\hypersetup{colorlinks=false,allbordercolors=blue,pdfborderstyle={/S/U/W 1}}
\usepackage[normalem]{ulem}

\begin{document}

\markboth{Daria Prokhorova, Shuzhe Shi, Evgeny Andronov}{Energy loss of heavy-flavor quarks in color string medium}

\catchline{}{}{}{}{}

\title{Energy loss of heavy-flavor quarks in color string medium}

\author{Daria Prokhorova}

\address{Department of Physics, Tsinghua University, Haidian  District\\
Beijing, 100084,
China\\
daria-prokhorova@mail.tsinghua.edu.cn}

\author{Shuzhe Shi}

\address{Department of Physics, Tsinghua University, Haidian  District\\
Beijing, 100084,
China\\
shuzhe-shi@tsinghua.edu.cn}

\author{Evgeny Andronov}

\address{Physics Department, St. Petersburg State University, Peterhof\\
St. Petersburg, 198504,
Russia\\
e.v.andronov@spbu.ru}

\maketitle

\begin{history}
\received{Day Month Year}
\revised{Day Month Year}
\accepted{Day Month Year}
\published{Day Month Year}
\end{history}

\begin{abstract}
The paper presents preliminary estimates of heavy-flavor (HF) quark energy loss during its propagation through the non-equilibrated medium formed in minimum bias proton-proton (p+p) collisions at LHC energies. The study is inspired by the ongoing hot debates on whether tiny droplets of Quark-Gluon Plasma can be created in collisions of small systems. In this work, we model a p+p event with a fluctuating number of color strings originated from multi-pomeron exchanges. Considered longitudinal oscillations of strings dynamically initialize medium at each time step. Their varying overlaps create fluctuations in the color field energy density that governs the elastic scattering rate of HF quarks with the gluons present within the string volume. We calculate the transverse momentum dependence of the momentum loss for charm (anti-)quarks that are produced in initial hard scatterings and traverse the described environment. The simulation is performed using a developed hybrid approach on an event-by-event basis. Our results show significantly lower HF quarks energy loss compared to that obtained in the expanding hydrodynamic scenario of the new EPOS4HQ model.
\end{abstract}


\ccode{PACS Number(s): 12.38.Mh, 12.39.-x, 14.65.Dw}

\section{Introduction}

For the past decades, significant experimental~\cite{Heinz:2000bk, BRAHMS:2004adc, PHOBOS:2004zne, PHENIX:2004vcz, STAR:2005gfr,ALICE:2010khr,ATLAS:2010isq,CMS:2011iwn,LHCb:2015coe,ALICE:2022wpn,CMS:2024krd}
and theoretical efforts~\cite{Cabibbo:1975ig, Karsch:2001cy, Bazavov:2011nk}
have been put into studying a unique state of matter called Quark-Gluon Plasma (QGP)~\cite{Shuryak:1977ut}. Quarks and gluons that are normally confined within hadrons become asymptotically free~\cite{Gross:1973id, Politzer:1973fx} at extreme temperature and energy density of QGP. This kind of hot and dense medium is also believed to fill the Universe during first few microseconds after the Big Bang. Somewhat similar evolution of the Little Bang is expected in relativistic heavy ion collisions (HICs)~\cite{Heinz:2013wva}. 

The ``standard model'' of HICs suggests~\cite{Bjorken:1982qr, Busza:2018rrf} that as Lorentz-contracted nuclei pass through each other, the initial stage is characterized by hard scatterings and severe energy stopping. During this phase, lasting approximately 1 fm/\textit{c}, the system is highly non-uniform and its evolution is governed by large pressure gradients. It is usually assumed that within this short period, the plasma reaches local thermal equilibrium, forming a droplet of near-perfect liquid that can be described by relativistic hydrodynamics~\cite{Romatschke:2017ejr}. However, the pre-equilibrium stage remains poorly constrained and provides only vague initial conditions for the consequent hydrodynamic evolution that is usually applied for up to 10 fm/\textit{c} to model the expansion and cooling down of QGP. Finally, the bulk undergoes a phase transition from deconfined state to a hadron gas and produced hadrons free-stream towards detectors.

Thus, determining whether QGP was created for a given system size and collision energy relies on studying final-state particles for a plethora of theoretically predicted signals~\cite{Harris:1996zx} that mark the onset of deconfinement. Among them are strangeness enhancement~\cite{Koch:1986ud}, quarkonium suppression~\cite{Matsui:1986dk}, azimuthal flows~\cite{Sorge:1996pc, Csernai:1999nf}, and jet quenching~\cite{Gyulassy:1990ye, Wang:1992qdg}. Analysis of such experimental observables can reveal the fundamental properties of the strongly interacting matter, e.g.~\cite{Heinz:2013th, Mishra:2022kre} the parameters of the phase transition, the viscosity and vorticity of QGP, its equation of state, etc.

Surprisingly, several signatures characteristic of QGP have been observed in matter produced in collisions of small systems, including high-multiplicity proton-proton (p+p) interactions~\cite{ALICE:2016fzo, CMS:2010ifv, ATLAS:2015hzw, ATLAS:2016yzd}. These intriguing observations scrutinize the commonly perceived simplicity of p+p collisions, suggesting that the matter produced here can also exhibit fluid-like behavior~\cite{Dusling:2015gta}. This fact questions the limitations on QGP formation in such environments~\cite{Romatschke:2016hle} and redefines the scope of applicability for hydrodynamic frameworks~\cite{Weller:2017tsr,Zhou:2020pai,Ambrus:2022qya}. Alternative explanations~\cite{Florkowski:2017olj} propose that the observed collectivity may instead arise from hydrodynamic attractors~\cite{Heller:2013fn}, the dynamics of strong initial-state color fields~\cite{Lappi:2006fp, Kovner:2010xk}, or momentum-space isotropization~\cite{Romatschke:2016hle}. 

To address the puzzle, we study the influence of the controversial short-lived medium, formed in p+p interactions, on a penetrating probe. Ideal candidates are heavy-flavor (HF) quarks ($c, \Bar{c}, b, \Bar{b})$ as they are predominantly~\cite{Dong:2019unq} produced in the primary hard scatterings and traverse the developing system. Being much heavier than the medium constituents, they avoid rapid thermalization and can retain information about the full space-time history of the fireball evolution~\cite{Svetitsky:1987gq,Moore:2004tg}. 
Through interactions with the different states of the background medium, HF quarks can lose energy, which suppresses the final-state spectra of HF hadrons. There are several theoretical frameworks that accounts for this, e.g. the propagation of HF quarks in pre-equilibrium Glasma fields~\cite{Schenke:2008gg, Carrington:2015xca, Mrowczynski:2017kso}, transport models used for its motion in a thermal medium phase~\cite{Cao:2018ews}, full-scale event generators~\cite{Zhao:2023ucp}, and multi-stage approaches~\cite{Singh:2025duj}.

This analysis is part of our ongoing systematic effort~\cite{Andronov:2023vnh, Prokhorova:2023hgq, Prokhorova:2024ext} to test the quark-gluon string model's description of collective effects observed in p+p interactions without assuming QGP formation. Within this approach, the initial state of the collision consists of parallel quark-gluon strings, formed between colliding partons, while the final state is obtained from string fragmentation into 
observable hadrons. The picture originates from the pre-QCD Regge--Gribov formalism~\cite{Gribov:1967vfb} that describes high-energy elastic scattering amplitudes in terms of multiple pomeron exchanges. With the advent of QCD, the dominant contribution of the QCD topological expansion in the large number of color, $N_c$, and large number of flavor, $N_f$, limits has been shown~\cite{Veneziano:1974ag,Veneziano:1974fa,Veneziano:1976wm} to be given by the cylindrical diagram representing each pomeron exchange. The unitarity cut of such diagram leads to its space-time localization: two rapidity chains are formed~\cite{Capella:1992yb, Werner:1993uh, Kaidalov:1982xg,Artru:1979ye} that can be viewed as color flux tubes and be effectively described by the Cornell potential~\cite{Eichten:1974af}.

The color string model of multi-particle production appeared to be effective both in phenomenological calculations, see e.g. dual parton model~\cite{Capella:1992yb}, string percolation model~\cite{Braun:1991dg}, and as the basis of many Monte-Carlo event generators, e.g. EPOS~\cite{Werner:2023zvo}, PYTHIA~\cite{Sjostrand:2019zhc}, HIJING~\cite{Wang:1991hta}, AMPT~\cite{Zhang:1999bd}, and PHSD~\cite{Cassing:2008sv}. Some of these frameworks consider interactions of strings~\cite{Andersson:1991er} (e.g. color reconnection, string ropes, string shoving etc.), which results in collective-like effects in the final state. Other employ a hybrid multistage picture, enabling the separation of initial- and final-state effects~\cite{Jafarpour:2025tsx}. In turn, our simpler approach accounts for color flux tubes finite transverse size~\cite{Cea:2014uja,Nishino:2019bzb}, which leads to their overlaps~\cite{Bierlich:2014xba} and consequent reorientation of the color fields within the fused string domains~\cite{Braun:1997ch,Bali:2000un}.

In this study, we model the propagation of charm quarks within string medium of fluctuating density and estimate its momentum loss from elastic scatterings off gluons confined within color tubes. The ultimate objective, however, is to compare model predictions with the p+p data~\cite{ALICE:2020wfu,ALICE:2021psx,ALICE:2021npz, CMS:2023frs} on flow and modification of $p_T$ spectra for hadrons containing HF quarks. Thus, a detailed treatment of the hadronization process in the current framework is left for a dedicated future work.

The paper is organized as follows. Section~\ref{sec.2} presents our developed hybrid approach. For brevity, we focus on a summary of the model~\cite{Andronov:2023vnh,Prokhorova:2023hgq,Prokhorova:2024ext} and highlight the specific modifications made to its workflow. Section~\ref{sec.3} reports our preliminary findings and includes a brief analysis. Summary and outlook Section closes the paper.

\section{Developed hybrid approach}\label{sec.2}
This work extends our previous Monte-Carlo model of interacting quark-gluon strings~\cite{Andronov:2023vnh, Prokhorova:2023hgq, Prokhorova:2024ext} that provided a thorough description of measured rapidity correlations and fluctuations, as well as of the ridge structure observed in inelastic p+p collisions at LHC energies. Here, we apply this approach to model the in-medium modifications of heavy-flavor quarks transverse momentum spectra. 

\subsection{Hard vs soft regimes}
In each event, following the QCD factorization theorem~\cite{CTEQ:1993hwr}, we treat hard and soft processes separately. The generation of HF quarks via $g+g \rightarrow Q\bar{Q}$ and $q+\bar{q} \rightarrow Q\bar{Q}$ channels (where $Q = c, b$ and $q = u, d, s$) is handled by the PYTHIA8.3 event generator~\cite{Bierlich:2022pfr}. In turn, the number of exchanged soft pomerons, $n_\mathrm{pom}$, as the function of the energy available for soft processes, $\sqrt{s}_\mathrm{soft}$, is given by the Regge-theory parametrization~\cite{Kaidalov:1982xe} that neglects the three-pomeron vertices
\begin{equation}\label{MPE}
      \footnotesize  
P\left(n_\mathrm{pom}\right)=C\left(z\right)\frac{1}{zn_\mathrm{{pom}}}\left(1-\exp\left(-z\right)\sum_{l=0}^{n_\mathrm{{pom}}-1}\frac{z^{l}}{l!}\right),\quad
z=\frac{2w\gamma s_\mathrm{soft}^{\Delta}}{R^{2}+\alpha^{\prime}\, \ln s_\mathrm{soft}}.
\end{equation}
Here, $C(z)$ is a normalizing coefficient,
$w = 1.5$ is the quasi-eikonal parameter related to the small-mass diffraction dissociation of incoming hadrons, $\Delta = \alpha(0) - 1 = 0.2$ is the residue of the pomeron trajectory, 
{$\alpha(0)$ is the intercept of the pomeron trajectory,} $\gamma = 1.035~\mathrm{GeV}^{-2}$ and $R^2 = 3.3~\mathrm{GeV}^{-2}$ characterize the coupling of the pomeron trajectory with the initial hadrons, $\alpha' = 0.05~ 
\mathrm{GeV}^{-2}$ is the slope of the pomeron trajectory (all values from Ref.~\refcite{Vechernin:2020piz}). Thus, extracted HF pairs are inserted into the non-perturbative environment consisting of $n_\mathrm{str} = 2n_\mathrm{pom}$ color string.

\subsection{Proton composition}
For a given event, the found value of $n_\mathrm{str}$ defines the number of partons (valence quarks, diquarks, and sea quark-antiquark pairs) within each of the two colliding protons. The longitudinal momentum of each parton is $p_z = x \cdot p_\mathrm{beam}$, where the beam momentum fraction $x$ is sampled from parton distribution functions (PDFs)~\cite{Lai:2010vv,Gao:2013xoa}, while transverse momentum $p_T = 0$. The sampling scale $Q^2$ is provided event-by-event by PYTHIA8.3. We ensure energy-momentum conservation for the partons within a proton using the procedure based on the algorithm~\cite{Prokhorova:2024ext}. 

\subsection{Strings formation}

The model initializes the system as $n_\mathrm{str}$ purely longitudinal strings: we randomly link partons from the colliding protons (each contains $n_\mathrm{str}$ partons), permitting the following combinations: $q+\bar{q}$, $\bar{q}+q$, $q+qq$ and $qq+q$ (where $q = u, d, s$, with no distinction between sea and valence quarks). If the target number of strings is not reached, the parton connections are first reshuffled. If this fails, the proton pair is regenerated.

For a parton with mass $m$ and initial longitudinal momentum $p^\mathrm{init}_z$, the initial rapidity, $y^\mathrm{init}$, is given by:    
\begin{equation}\label{init_rapidity}
    y^\mathrm{init} = \sinh^{-1}\left(\frac{p^\mathrm{init}_z}{m}\right),
\end{equation}
which, calculated for both ends of each string, defines the initial event configuration of color flux tubes in the rapidity space. In the transverse plane, the strings centers coordinates are sampled according to the Gaussian distribution centered at the origin with a width of 0.5 fm. 

\subsection{Strings evolution}
The dynamical evolution of strings in longitudinal space~\cite{Shen:2017bsr} is governed by a string tension $\sigma = 1$~GeV/fm which acts on the partons at the string ends as
\begin{equation}\label{part_EOMs}
    \frac{dp_z}{dt} = \pm\sigma, \qquad \frac{dE}{dz} = \pm\sigma,
\end{equation}
where $t$ and $z$ are time-space coordinates of a parton and $E = \sqrt{m^2+p_z^2}$ is its energy. The periodic yo-yo motion of the string is implemented by defining a maximum proper time $\Delta\tau_\mathrm{max}$ for each parton, after which its longitudinal momentum $p_z$ vanishes and the sign of $\sigma$ in Eq.~(\ref{part_EOMs}) flips, as
\begin{equation}\label{taumax}
    \Delta\tau_\mathrm{max} = \frac{m}{\sigma} \sqrt{2\left( \mathrm{cosh}(\Tilde{y}^\mathrm{init} - 1)\right)},
\end{equation}
where $\Tilde{y}_\mathrm{init}$ is partons' initial rapidity in the string rest frame. Note that $\Delta\tau_\mathrm{max}$ is defined independently for the left and right string ends, as they may have different masses and initial momenta.

To maintain a consistent spacetime picture for the evolution of all strings in an event and the propagation of HF quarks through the string medium, the lab frame is used as the common reference frame. Thus, the relation between the partons' proper time $\Delta\tau$ and the lab time $\Delta t$ intervals~\cite{Shen:2017bsr} is given by
\begin{equation}\label{dtdtau}
     \Delta t = \Delta\tau\left( \pm \frac{\sigma\Delta\tau}{2m}\mathrm{sinh}(y^\mathrm{init}) + \mathrm{cosh}(y^\mathrm{init})\sqrt{\frac{\sigma^2\Delta\tau^2}{4m^2}+1}\right).
\end{equation}
Due to the periodic sign change of $\sigma$ in Eqs.~(\ref{part_EOMs}) and (\ref{dtdtau}), $\Delta\tau(\Delta t)$ dependence must be defined piecewise with respect to $\Delta\tau_\mathrm{max}$.  The resulting $\Delta \tau^{*}$, that is free of half periodic contributions, is then used to calculate the total rapidity shift $y^\mathrm{shift}$ of the parton after a lab frame time interval $\Delta t$ as
\begin{equation}
    y^\mathrm{shift} = \cosh^{-1}\left( \frac{\sigma^2(\Delta\tau^{*})^2}{2m^2} + 1 \right).
\end{equation}
The final rapidity of a string end, $y^\mathrm{fin}$, is determined as
$y^\mathrm{fin} = \pm y^\mathrm{init} \pm y^\mathrm{shift}$, where $\pm1$ coefficients are independent and are determined from the sign of $y^\mathrm{init}$ and the parity of $\Delta \tau^{*}$, respectively. In this work, we neglect the evolution of string density in the transverse plane, which was considered in Ref.~\refcite{Prokhorova:2024ext}.

\subsection{String medium characteristics}\label{subsec2.5}

Due to finite transverse size, chosen as $r_\mathrm{str} = 0.25~\mathrm{fm}$, strings can overlap, which results in a percolation picture in the transverse plane. Together with combined fluctuations in string length and rapidity positions, it creates a highly inhomogeneous three-dimensional (3D) string density distribution. To account for this, we introduce a fine binning of the mixed configuration-momentum space into cells of chosen size $dX = dY = 0.05$~fm and $dy = 0.1$, defined by transverse coordinates ($X$, $Y$) and rapidity ($y$).

The superposition of color flux tubes reorients the color fields, as additional color charges are introduced in the string system~\cite{Bali:2000un}. Consequently, the resulting energy density in a cell, $\varepsilon_\mathrm{cell}$, covered by $k_\mathrm{cell}$ strings~\cite{Braun:1997ch} is
\begin{equation}\label{energydensity}
    \varepsilon_\mathrm{cell} = \frac{\sigma\sqrt{k_\mathrm{cell}}}{\Delta S_\mathrm{str}},
\end{equation}
where $\Delta S_\mathrm{str} = \pi r^2_\mathrm{str}$ is the transverse area of a string not varying along the rapidity direction, albeit the picture may be more complex~\cite{Luscher:1980iy}. Therefore, the medium energy density distribution is discrete, as it depends on the integer number of strings per cell, $k_\mathrm{cell} \in \mathbb{N}$. Simple estimates yield an energy density range of $5 < \varepsilon_\mathrm{cell} < 50$~$\mathrm{GeV}/\mathrm{fm^3}$ or 
$0.04 < \varepsilon_\mathrm{cell} < 0.4$ ~$\mathrm{GeV}^4$ for $1 < k < 100$.

Since the model accounts for rapidity evolution of each string, we recalculate cell energy densities at each time step. For simplicity, we still denote it as $\varepsilon_\mathrm{cell}$, which remains a function of time.

\subsection{Propagation of HF quarks}

This work simulates system evolution up to $t_\mathrm{max} = 1.5$~fm/\textit{c}, chosen as the approximate global pre-hadronization timescale~\cite{Andersson:1983ia}. The dynamics is computed with a temporal resolution of $\Delta t = 0.1$~fm/\textit{c}. Within each time step, the string medium is updated first, and subsequently, the quark propagation is calculated in the frozen string configuration. Our analysis focuses on the midrapidity region, $|y|<0.5$, to compare simulated HF hadron flow with data~\cite{ALICE:2020wfu,ALICE:2021psx,ALICE:2021npz, CMS:2023frs} in future studies.

In an event, the four-momenta of HF quarks, generated by hard processes in PYTHIA8.3, serve as input for their subsequent propagation through the medium. Each pair is assigned a common transverse position ($X$-$Y$), sampled from a distribution proportional to the 3D string density across the ensemble of medium cells. 

HF quark propagation through the frozen medium is modeled via deterministic collisional energy-loss approach~\cite{Braaten:1991jj}. A quark follows straight-line trajectory, perpendicular to the strings and defined by its initial transverse momentum $p_T$. This confines quark motion within its original rapidity slice, neglecting stochastic interactions and large-angle scatterings.
 
\subsection{In-medium transverse momentum loss of HF quarks}

There exist several treatments of the in-medium interactions of HF quarks. In the Color-Glass condensate framework, HF quark interactions with classical Yang-Mills fields are described by collision terms of a Fokker-Planck equation that capture Glasma dynamics through field correlators~\cite{Carrington:2020sww,Carrington:2021dvw}. A recent multi-stage approach~\cite{Singh:2025duj} goes further by considering the Brownian motion of charm quarks in this pre-equilibrium phase with MARTINI event generator~\cite{Schenke:2009gb} and quantifying their interactions with the medium via drag and diffusion coefficients. In the EPOS4HQ~\cite{Zhao:2023ucp} version of the EPOS4 model~\cite{Werner:2023zvo},
HF quarks experience both elastic~\cite{Gossiaux:2008jv} and radiative~\cite{Aichelin:2013mra} collisions with the thermal partons of expanding QGP, formed in spatial regions where the energy density is above a critical value. 

This study proposes a complementary bottom-up approach that computes the medium's energy density profile from a 3D ensemble of gluon-populated color flux tubes. The corresponding local effective temperature determines the elastic collision rate between HF quarks and gluons. The resulting HF energy loss accumulates only during this initial pre-equilibrium phase and is isolated from hydrodynamic effects, as our model excludes subsequent evolution.

To find the relevant scattering cross-sections, we adopt the approach from CUJET3 framework~\cite{Xu:2014tda} built on a non-perturbative microscopic model for
hot medium as semi-quark-gluon-monopole plasma. It employs the Thoma--Gyulassy elastic cross-section~\cite{Thoma:1990fm, Peigne:2008nd} for a collision of partons $i$ and $j$, based on Bjorken’s estimation of elastic energy loss in QGP~\cite{Bjorken:1982tu},
\begin{equation}\label{eq_elastic_xsection}
\frac{\mathrm{d}\sigma_{i,j}}{\mathrm{d}\hat{t}} = \frac{2\pi \alpha_s^{2}}{\hat{t}^2} c_{i,j},
\end{equation}
where $\hat{t} = -q^2$ is the momentum transfer, $\alpha_s$ is the coupling constant, the color factor, $c_{i,j}$, equals to $4/9$, $1$, and $9/4$ for $\{i,j\}=\{q,q\}, \{q,g\}, \{g,g\}$ respectively, where $q$ stands for quark and $g$ for gluon. In our case of quark-gluon interactions, $c_{i,j} = 1$.

In CUJET3/CIBJET~\cite{Xu:2014tda, Shi:2018lsf} frameworks, the following thermal running coupling form, inspired by lattice data~\cite{Liao:2008jg}, is employed:
\begin{equation}
\alpha_s^{\mathrm{th}} (Q^2) = \frac{\alpha_c}{1+\frac{(11-\frac{2}{3}n_\mathrm{f})}{4\pi}\alpha_c\ln(Q^2/T_c^2)},
\label{eq.runningcouplingthermal}
\end{equation}
where $n_\mathrm{f}$ is the number of flavors, $\alpha_c$ is the maximum value of $\alpha_s^{\mathrm{th}} (Q^2)$ at the non-perturbative scale $T_c$. The value $\alpha_c = 0.9$ at $T_c = 0.16$ GeV was determined from comprehensive global $\chi^2$ analysis of nuclear collision data at RHIC and LHC in Ref.~\refcite{Shi:2018izg}. 

In thermal medium, a screening phenomenon arises~\cite{Gossiaux:2008jv,Gossiaux:2009hr} due to the gluon thermal Debye mass, $\mu$, determined from solving a self-consistent equation 
\begin{equation}
\mu = \sqrt{4\pi\alpha^{\mathrm{th}}_s(\mu^2)}\cdot T\cdot \sqrt{1+n_\mathrm{f}/6},
\end{equation}
where $T$ is the medium temperature. Strictly speaking, both $T$ and $\mu$ are functions of the transverse coordinate $\bold{z} = (x_0+t\cos\phi,y_0 + t\sin\phi,t)$, determined by HF quark's initial position $(x_0,y_0)$ and propagation time $t$, according to the notations in Ref.~\refcite{Xu:2014ica}, which accounts for generic space-time dependent plasma geometries. For simplicity, we omit the explicit notations $T(\mathbf{z})$ and $\mu(\mathbf{z})$ but will incorporate their spatial dependence via the subscript $_{\mathrm{cell}}$ to reflect the binned structure of the medium.

Thus, the infrared finite form of the elastic cross-section from Eq.~(\ref{eq_elastic_xsection}) reads 
\begin{equation}
\frac{\mathrm{d}\sigma_{i,j}}{\mathrm{d}\hat{t}} = \frac{2\pi \alpha_s^{2}}{(\hat{t}+\mu^2)\hat{t}} c_{i,j}
\end{equation}
and the momentum loss rate, $\mathrm{d}p_i/\mathrm{d}t$, of a parton of type $i$ scattered off partons of types $j$ is generally defined as
\begin{equation}\label{full_eq}
\frac{\mathrm{d}p_i}{\mathrm{d}t} = \sum_j\int \frac{\mathrm{d}^3k}{(2\pi)^3}\; \rho_j(k)\; \Phi \int_{\hat{t}_\mathrm{min}}^{\hat{t}_\mathrm{max}} \mathrm{d}\hat{t} \frac{\mathrm{d}\sigma_{i,j}}{\mathrm{d}\hat{t}}\;\Delta p_i(\hat{t},k,p_i), 
\end{equation}
where $\rho_j(k)$ is the momentum distribution of medium constituents of type $j$, $\Phi$ is the incoming probe flux, $\Delta p_i(\hat{t},k,p_i)$ is the one-collision quark's transverse momentum loss at given $\hat{t}$, $k$, and $p_i$. Momentum transferred ranges from $\hat{t}_\mathrm{min}$ to $\hat{t}_\mathrm{max}$, which is the approximation to regulate the infrared and ultraviolet divergences. 

\subsubsection{Kinematics}
For the elastic scattering $Q(p) + g(k) \rightarrow Q(p') + g(k')$ of a HF quark on a massless gluon, the momentum transfer is $q = p - p' = k' - k$. Energy-momentum conservation and the on-shell conditions imply $\hat{t} \equiv -q^2 = 2k \cdot k' = 2k \cdot (p-p')$. 
The resulting change in the massive parton's momentum is
\begin{align}
\begin{split}
    \Delta p = p - p'\cos\theta_{pp'} \approx& \frac{\hat{t}}{2k} 
    \Big(\frac{E(p-E\cos\theta)}{(E-p\cos\theta)^2} - \frac{k\cos\theta}{E-p\cos\theta}\Big),
\end{split}
\end{align}
where $E = \sqrt{p^2 + m^2}$ and $\theta$ is the angle between momentum of the HF quark and medium gluons. 
It shall be worth noting that CUJET3~\cite{Xu:2014ica} further takes the ultra-relativistic approximation that $\Delta p\approx\frac{\hat{t}}{2k(1-\cos\theta)}$, and such approximation has been omitted here since our study extends beyond hard probes. The flux factor $\Phi$, which accounts for the relative orientation between the target and projectile, is given by $1-\cos\theta$. Thus, for a running strong coupling from Eq.~\eqref{eq.runningcouplingthermal}, one finds
\begin{align}
\begin{split}
\int_{\hat{t}_\mathrm{min}}^{\hat{t}_\mathrm{max}} \mathrm{d}\hat{t} \frac{\mathrm{d}\sigma_{i,j}}{\mathrm{d}\hat{t}}\; (p-p'\cos\theta_{pp'})
=&
    -\frac{\pi c_{i,j} \alpha^{\mathrm{th}}_s(\hat{t}_\mathrm{max}) \alpha^{\mathrm{th}}_s(\hat{t}_\mathrm{min})}{k} \ln\left(\frac{\hat{t}_\mathrm{max}}{\hat{t}_\mathrm{min}}\right)\\
    &
    \times\Big(\frac{E(p-E\cos\theta)}{(E-p\cos\theta)^2} - \frac{k\cos\theta}{E-p\cos\theta}\Big).
\end{split}
\end{align}
Here, the momentum transferred is limited from below by $\mu^2$ and from above by 
\begin{align}
\begin{split}
\hat{t}_\mathrm{max} = \frac{4(k\cdot p)^2}{2 k\cdot p + m^2}&\approx 2Ek - m^2 + \frac{m^4}{4kp}\ln\frac{m^2+2k(E+p)}{m^2+2k(E-p)}\\ 
&\approx 6ET - m^2 + \frac{m^4}{12Tp}\ln\frac{m^2+6T(E+p)}{m^2+6T(E-p)}
\end{split}
\end{align} 
to screen the otherwise ultraviolet divergent logarithm. For the massless assumption in CUJET3~\cite{Xu:2014ica}, $\hat{t}_\mathrm{max}\approx 2kp\approx2\langle k\rangle p\approx6Tp$. 

Thus, the momentum loss rate for the considered elastic scattering of HF quark off medium gluons reads
\begin{align}\label{eq.dpdt_general}
\begin{split}
\frac{\mathrm{d}p}{\mathrm{d}t} =& \pi\alpha^{\mathrm{th}}_s(\hat{t}_\mathrm{max}) \alpha^{\mathrm{th}}_s(\hat{t}_\mathrm{min}) \ln\left(\frac{\hat{t}_\mathrm{max}}{\hat{t}_\mathrm{min}}\right) \int \frac{\mathrm{d}^3k}{(2\pi)^3}\; \frac{\rho(k)}{k}\; \\&
    \times\Big(\frac{E(p-E\cos\theta)}{(E-p\cos\theta)^2} - \frac{k\cos\theta}{E-p\cos\theta}\Big)(1-\cos\theta),
\end{split}
\end{align}
and should be solved recursively with the initial condition $p|_{t=0} = p^\mathrm{init}$ up to a cutoff time $t_\mathrm{max}$, or until the quark's momentum drops below zero. We again omit the coordinate dependence of $p(\bold{z})$ and $ E(\bold{z})$ since the Monte-Carlo algorithm treats the propagation in discrete steps by computing the momentum loss in each medium cell and updating quark's momentum and energy, $p$ and $E$, for the next iteration. 

To evaluate Eq.~\eqref{eq.dpdt_general}, one needs to make an assumption about the momentum distribution of medium constituents, $\rho(k)$. 

\subsubsection{Ideal Bose gas of gluons}
As a starting point, we assume a simple parameterization for the gluonic degrees of freedom inside color flux tubes using the ideal Bose gas momentum distribution~\cite{Kapusta:2006pm}
\begin{equation}\label{eq.BoseStatisitc}
\rho(k) = \frac{g_\mathrm{gl}}{e^{k/T^\mathrm{eff}}-1},
\end{equation}
where $g_\mathrm{gl} = N_\mathrm{pol}\cdot (N^2_c - 1) = 16$ is the degeneracy factor for gluons calculated from number of polarizations, $N_\mathrm{pol} = 2$, and number of colors, $N_c = 3$. This form of $\rho(k)$ imposes an effective thermal description by assigning the string medium an effective temperature, $T^\mathrm{eff}$, derived from its energy density. This allows us to directly compare the energy loss of a heavy-flavor quark propagating in our string environment to the energy loss it would experience in a thermally equilibrated pure-gluon QGP of identical energy density.

Thus, evaluating the integral in Eq.~\eqref{eq.dpdt_general} in spherical coordinates yields
\begin{align}\label{eq.loss_ideal_bose_gluons}
\begin{split}
\frac{dp}{dt} & = \pi C_R [T^\mathrm{eff}_\mathrm{cell}]^2\alpha^{\mathrm{th}}_s(\mu^2_\mathrm{cell})\alpha^{\mathrm{th}}_s(6ET^\mathrm{eff}_\mathrm{cell})\\
&\times\mathrm{ln}\bigg[\frac{\bigg(6ET^\mathrm{eff}_\mathrm{cell} - m^2 + \frac{m^4}{12T^\mathrm{eff}_\mathrm{cell}p}\ln\frac{m^2+6T^\mathrm{eff}_\mathrm{cell}(E+p)}{m^2+6T^\mathrm{eff}_\mathrm{cell}(E-p)}\bigg)^{1/2}}{\mu_\mathrm{cell}}\bigg]\\
& \times \bigg\{\frac{E}{p^3}\Big((2E-p)p - (E^2+m^2-Ep)\mathrm{sinh^{-1}}\left(\frac{p}{m}\right)\Big)\\
& +\frac{12\zeta(3) T^\mathrm{eff}_\mathrm{cell}}{\pi^2}\frac{(E-p)}{p^3}\Big(E\,\mathrm{sinh^{-1}}\left(\frac{p}{m}\right)-p\Big)
\bigg\},
\end{split}
\end{align}
where $C_R = 4/3$ for quark probe, $\zeta(3)$ is the Riemann zeta function calculated at argument equal to three, running coupling prescription follows Ref.~\refcite{Peigne:2008nd} and logarithm argument is extended from one found in Ref.~\refcite{Xu:2014ica} by taking into account finite HF quark mass. To account for the possible transverse expansion of the medium, as in Ref.~\refcite{Shi:2018izg}, the energy of the HF quark in above equation should be replaced as $E \to \Gamma_\mathrm{cell} E$, where the correction factor $\Gamma_\mathrm{cell} = u^\mu_\mathrm{cell} n_\mu$ is defined by the cell four-velocity, $u^\mu_\mathrm{cell}$, and the quark's direction of motion, $n^\mu \equiv (1, \boldsymbol{\beta})$. However, this correction is beyond the scope of the current study.

\subsubsection{Anisotropic Bose gas of gluons}

To go beyond the assumption of an ideal thermalized medium and test the sensitivity of our results to the momentum distribution of medium constituents, we employ the Romatschke--Strickland (RS) parameterization of $\rho(k)$ from anisotropic hydrodynamics~\cite{Romatschke:2003ms,Romatschke:2004jh}. This approach models the momentum-space deformation of a non-equilibrium plasma, quantified by a parameter, $\xi > -1$, that governs the difference between the longitudinal ($k_z$) and transverse ($k_T$) components of the medium particles' momentum distributions, $f_\mathrm{RS}(k)$, as
\begin{equation}
    f_\mathrm{RS}(k) = e^{-\frac{\sqrt{k_T^2+(1+\xi)k_z^2}}{\Lambda}}.
\end{equation}
Here, $\Lambda$ is a characteristic momentum scale (distinct from a temperature, as the system is not thermalized). The parameter $\xi$ defines the shape of the distribution:
$\xi > 0$ corresponds to a contraction along the $k_z$ direction, while $-1 < \xi < 0$ provides a stretching of the distribution in $k_z$ direction. Although this form is mathematically equivalent to a simple rescaling of one momentum coordinate, it leads to profound insights into the non-equilibrium behavior and evolution of QGP~\cite{Romatschke:2016hle}. Moreover, viscous anisotropic hydrodynamics significantly extends the applicability of macroscopic hydrodynamic approaches, making it suitable for describing the evolution of small systems~\cite{Peng:2025gbj}.

In the context of our model, we apply this anisotropic momentum deformation to the Bose-Einstein distribution function
\begin{equation}\label{eq.AsymBoseStatisitc}
\rho(k) = \frac{g_\mathrm{gl}}{e^{\frac{\sqrt{k_T^2+(1+\xi)k_z^2}}{\Lambda}}-1}.
\end{equation}
Thus, evaluating the integral in Eq.~\eqref{eq.dpdt_general} in cylindrical coordinates for $\xi >0$ yields
\begin{align}\label{eq.loss_asym_bose_gluons}
\begin{split}
\frac{\mathrm{d}p}{\mathrm{d}t}
=& \;\pi C_\mathrm{R}\alpha^{\mathrm{th}}_s(\mu_\mathrm{cell}^2) \alpha^{\mathrm{th}}_s(6E\Lambda_\mathrm{cell})\\
&\times\ln\Big[\frac{\sqrt{6E\Lambda_\mathrm{cell} - m^2 + \frac{m^4}{12\Lambda_\mathrm{cell}p}\ln\frac{m^2+6\Lambda_\mathrm{cell}(E+p)}{m^2+6\Lambda_\mathrm{cell}(E-p)}}}{\mu_\mathrm{cell}} 
\Big]\\
& \times [\Lambda_\mathrm{cell}]^2(1-\varepsilon)\Bigg[
\frac{1}{\xi+\varepsilon^2}\cdot \bigg\{\frac{\mathrm{arctan} \sqrt{\xi}}{\sqrt{\xi}}\bigg(\xi+\frac{1+\xi}{1-\varepsilon}-\frac{2\xi}{\gamma}\bigg) \\
&+1+\mathrm{ln}\frac{1-\varepsilon}{1+\varepsilon}\cdot \left(\frac{\varepsilon}{\gamma} +\frac{1}{2}\right)\bigg\}+\frac{12\zeta(3)}{\pi^2}\cdot \frac{\Lambda_\mathrm{cell}}{E}\cdot \frac{1}{(\xi+1)^{\frac{3}{2}}}\\
&\times \bigg\{\frac{\mathrm{arctanh\sqrt{\gamma}}}{\gamma^\frac{3}{2}} - \frac{1}{\gamma }\bigg\}\Bigg],
\end{split}
\end{align}
where $\varepsilon = p/E$ and $\gamma = (\xi+\varepsilon^2)/(\xi+1)$. Performing a Taylor expansion around $\xi = 0$ and assuming $\Lambda_\mathrm{cell} = T_\mathrm{cell}$ recovers the result of Eq.~\eqref{eq.loss_ideal_bose_gluons}, but this approximation is contingent on the smallness of the parameter $\varepsilon$ as they both enter $\gamma$.

\subsubsection{Matching geometrical and computed medium energy density profiles.}
In this work, the medium is defined by the configuration of color strings, as discussed in Sec.~\ref{subsec2.5}. Consequently, the energy density $\varepsilon$ derived from the geometric formula in Eq.~(\ref{energydensity}) must be consistent with the value calculated from the chosen momentum distribution $\rho(k)$ in Eqs.~\eqref{eq.BoseStatisitc} and~\eqref{eq.AsymBoseStatisitc} for each medium cell.

For a thermal Bose gas of gluons, the energy density $\varepsilon$ depends on the temperature $T$ as follows:
\begin{align}\label{energy_dens_sym}
\begin{split}
\varepsilon (T) =&g_\mathrm{gl}\int_{0}^{\infty}\frac{d^3k}{(2\pi)^3} \frac{k}{e^{\frac{k}{T}}-1} = \frac{16\pi^2 }{30}T^4,
\end{split}
\end{align}
Thus, the effective temperature for string-based medium cell is given by
\begin{equation}\label{temp_sym}
    T^\mathrm{eff}_\mathrm{cell} = \sqrt[4]{\frac{\varepsilon_\mathrm{cell}}{16\pi^2/30}}.
\end{equation}
For the anisotropic Bose gas of gluons, the energy density $\varepsilon$ depends on both the momentum scale $\Lambda$ and the anisotropy parameter $\xi$ according to:
\begin{align}\label{energy_dens_asym}
\begin{split}
\varepsilon(\Lambda, \xi)=&g_\mathrm{gl}\int_{0}^{\infty}\frac{d^3k}{(2\pi)^3} \frac{k}{e^{\frac{\sqrt{k_T^2+(\xi+1)k_z^2}}{\Lambda}}-1} = \frac{16\pi^2}{60}\Lambda^4\left(\frac{1}{1+\xi}+\frac{\mathrm{arctan}\sqrt{\xi}}{\sqrt{\xi}}\right),
\end{split}
\end{align}
which converts to the result in Eq.~(\ref{energy_dens_sym}) with the Taylor expansion around $\xi = 0$. Therefore, the characteristic momentum scale for string-based medium cell is
\begin{equation}
\Lambda_\mathrm{cell} = \sqrt[4]{\frac{\varepsilon_\mathrm{cell}}{16\pi^2/60}}\cdot \left(\frac{1}{1+\xi}+\frac{\mathrm{arctan}\sqrt{\xi}}{\sqrt{\xi}}\right)^{-\frac{1}{4}}.
\end{equation}

\section{Preliminary results}\label{sec.3}

To verify the model workflow and obtain initial estimates, we first consider a simplified scenario and perform a brick test: a static homogeneous medium composed of $N$ fully overlapped strings, centered at $X=Y = 0$ and infinite in rapidity. Consequently, each medium cell has identical, time-independent properties $k_\mathrm{cell}$, $\varepsilon_\mathrm{cell}$, $T_\mathrm{cell}$, and $\mu_\mathrm{cell}$. HF quarks are initialized at $X=Y=0$ position to ensure they remain within the string environment throughout the propagation. It is further guaranteed  by artificially increasing the medium's transverse radius from $r^* \approx 1$~fm to $10$~fm, which requires a significant scaling of $N$ to maintain cell energy densities, $\varepsilon_\mathrm{cell}$, Eq.~\ref{energydensity}, comparable to the full dynamic simulation, where $ 1 < k_\mathrm{cell} < 100$. However, this $N$ adjustment demands minimal computational cost in a static medium.

\begin{figure}[th]
\centerline{\includegraphics[width=0.8\linewidth]{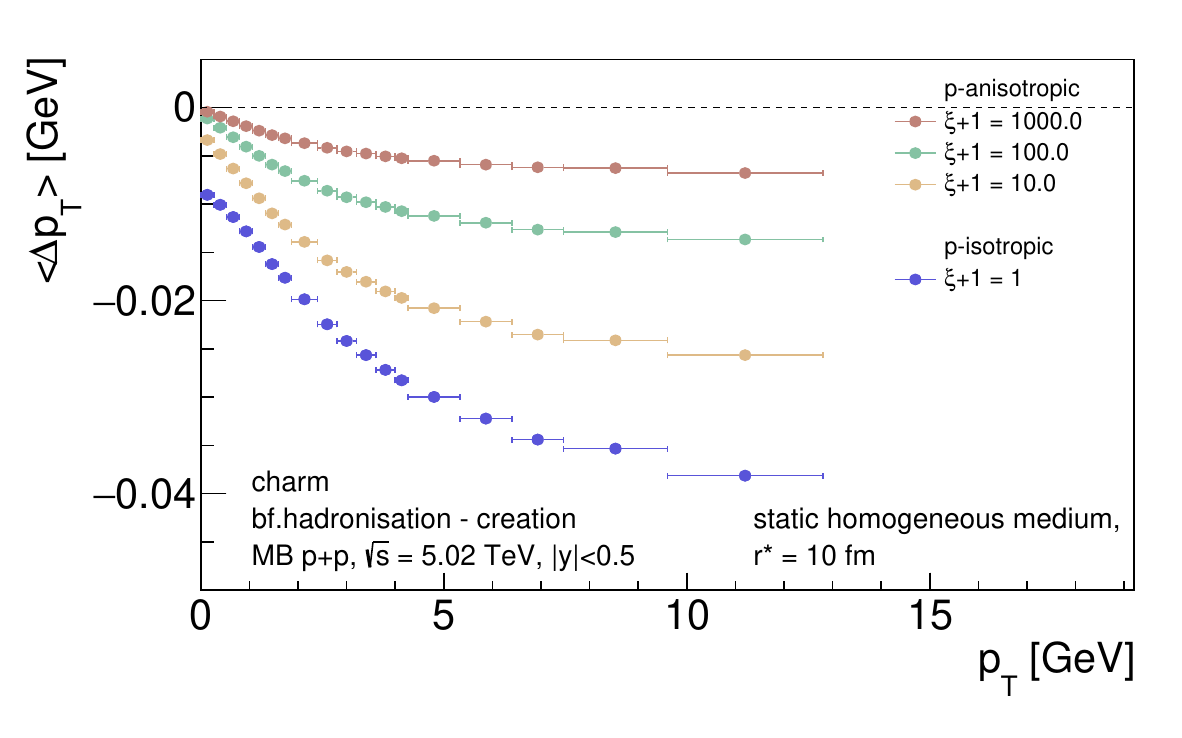}}
\caption{Transverse momentum loss of a charm quark as a function of its initial transverse momentum, calculated using a simplified model for minimum bias p+p interactions at $\sqrt{s} = 5.02$ TeV. The quark propagates for $t_\mathrm{max} = 1.5$~fm/\textit{c} through a static, homogeneous medium of radius  $r^*=10$~fm modeled by $N = 10^6$ fully overlapped quark-gluon strings. Results are shown for medium gluon momentum distributions following Bose statistics with different levels of 
$k_z-k_T$ anisotropy, parameterized by $\xi+1 = 1000, 100, 10, 1$.} 
\label{fig.simple_loss_variation_of_xi}
\end{figure}

\begin{figure}[th]
\centerline{\includegraphics[width=0.8\linewidth]{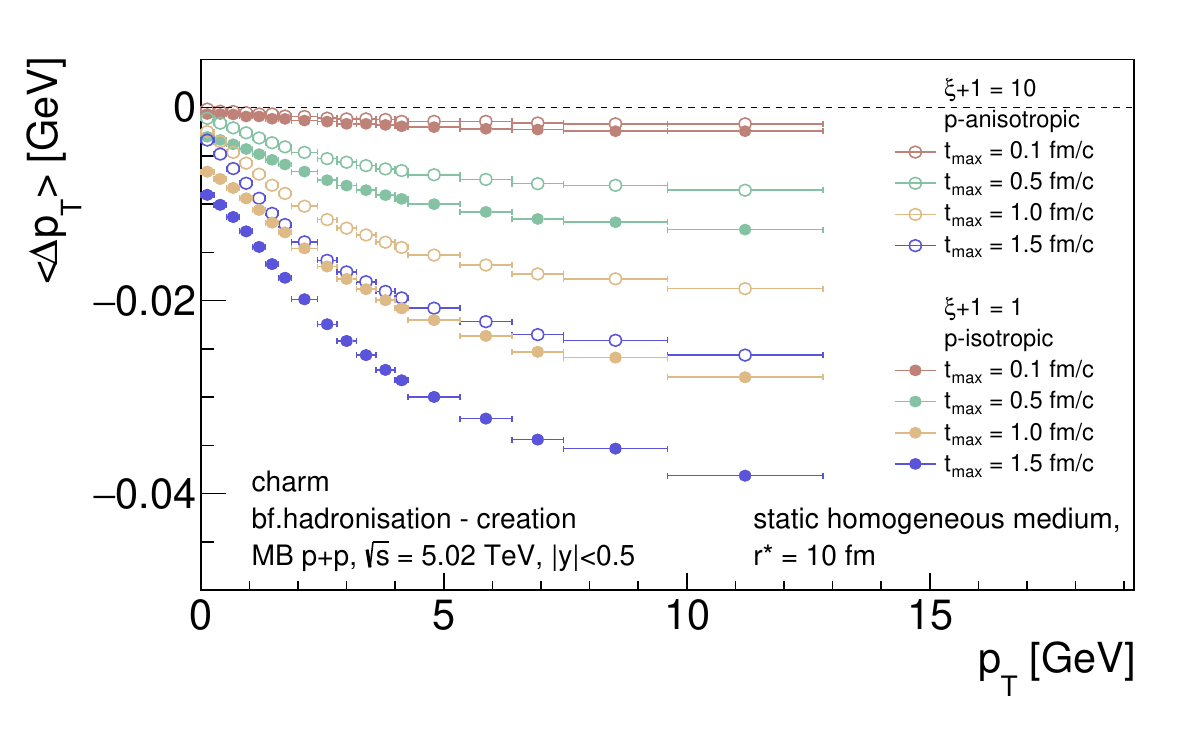}}
\caption{Transverse momentum loss of a charm quark as a function of its initial transverse momentum, calculated using a simplified model for minimum bias p+p interactions at $\sqrt{s} = 5.02$ TeV. The quark propagates through a static, homogeneous medium of radius $r^*=10$~fm, modeled by $N = 10^6$  fully overlapped quark-gluon strings, for different times $t_\mathrm{max} = 0.1, 0.5, 1.0, 1.5$~fm/\textit{c}. Results are shown for medium gluon momentum distributions following Bose statistics with $k_z-k_T$ anisotropy parameters $\xi+1 = 10, 1$.} 
\label{fig.simple_loss_variation_of_time}
\end{figure}

\begin{figure}[th]
\centerline{\includegraphics[width=0.8\linewidth]{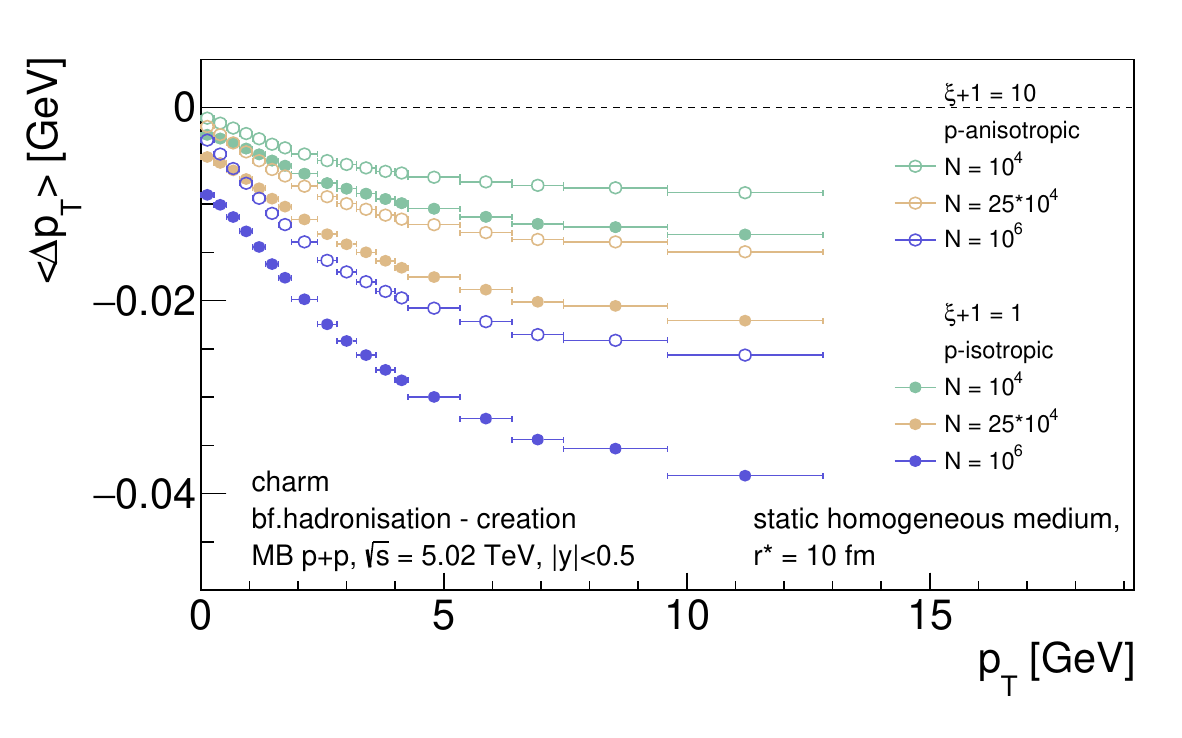}}
\caption{Transverse momentum loss of a charm quark as a function of its initial transverse momentum, calculated using a simplified model for minimum bias p+p interactions at $\sqrt{s} = 5.02$ TeV. The quark propagates for $t_\mathrm{max} = 1.5$~fm/\textit{c} through a static, homogeneous medium of radius $r^*=10$~fm, modeled by different numbers of fully overlapped quark-gluon strings $N = 10^4, 25\cdot10^4, 10^6$. Results are shown for medium gluon momentum distributions following Bose statistics with $k_z-k_T$ anisotropy parameters $\xi+1 = 10, 1$.} 
\label{fig.simple_loss_variation_of_N}
\end{figure}

We computed charm quark's momentum loss as a function of its initial momentum using our full-scale C++ framework for various model parameters ($\xi$, $t_\mathrm{max}$, $N$). The results, shown in Figs.~\ref{fig.simple_loss_variation_of_xi}--~\ref{fig.simple_loss_variation_of_N}, consistently agree with direct numerical integration  over propagation time in Mathematica of the loss rates, Eqs.~\eqref{eq.loss_ideal_bose_gluons} and~\eqref{eq.loss_asym_bose_gluons}.

Fig.~\ref{fig.simple_loss_variation_of_xi} demonstrates that increasing the momentum-space anisotropy of the medium gluons (larger $\xi + 1$ values) reduces the momentum loss of HF quarks over a propagation time of $t_\mathrm{max} = 1.5$~fm/\textit{c}, shown by the plots from dark blue to pale red. 
While $\xi + 1 = 100$ and $\xi + 1 = 1000$ represent extreme anisotropy, we focus on the moderate $\xi+1 = 10$ value for more detailed comparison with the isotropic $\xi+1 = 1$ scenario. Fig.~\ref{fig.simple_loss_variation_of_time} shows that different combinations of propagation time and anisotropy can yield similar quark damping. This degeneracy makes the momentum loss ambiguous unless both $t_\mathrm{max}$ and $\xi$ are known. 
The results for varying string densities ($N = 10^4, 25\cdot 10^4, 10^6$), presented in Fig.~\ref{fig.simple_loss_variation_of_N}, lead to a similar conclusion: a decrease in string density reduces the loss and complicates the distinction between the loss functions of the isotropic and anisotropic media.

\begin{figure}[th]
\centerline{\includegraphics[width=0.8\linewidth]{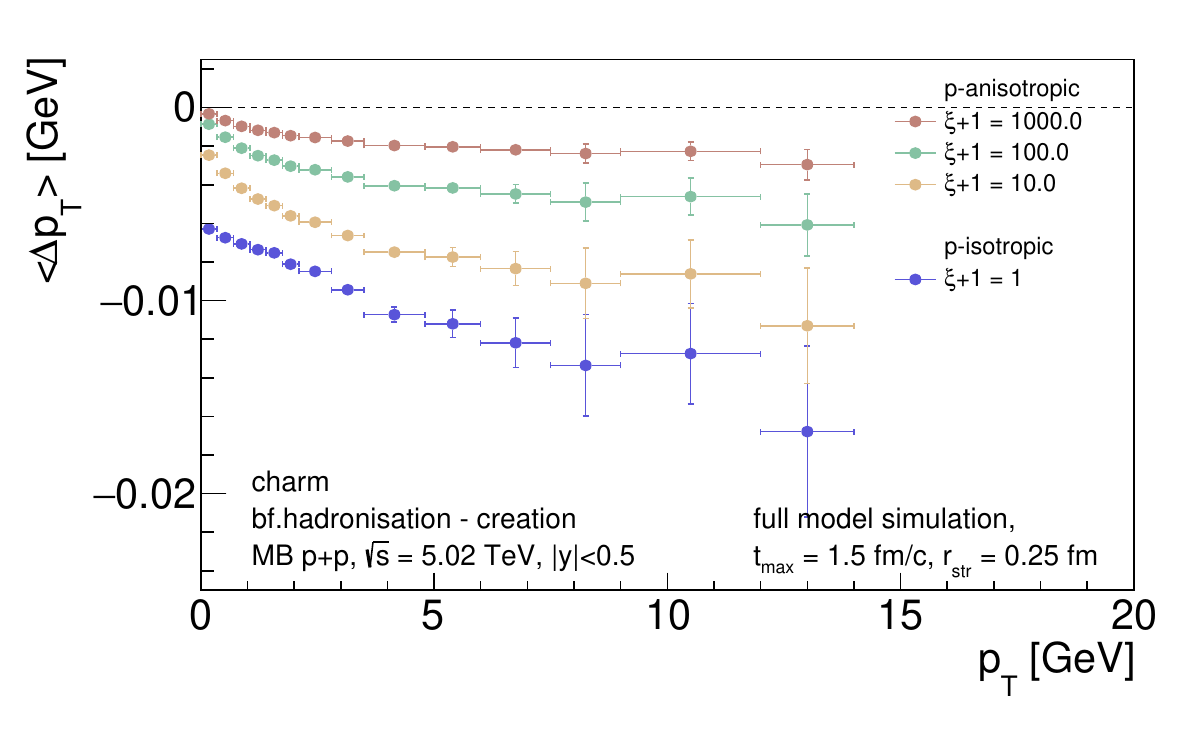}}
\caption{Transverse momentum loss of a charm quark as a function of its initial transverse momentum, calculated using the full dynamic model for minimum bias p+p interactions at $\sqrt{s} = 5.02$ TeV. Quark propagates for $t_\mathrm{max} = 1.5$~fm/\textit{c} through evolving in rapidity inhomogeneous medium of overlapped quark-gluon strings. Results are shown for medium gluon momentum distributions following Bose statistics with different levels of 
$k_z-k_T$ anisotropy, parameterized by $\xi+1 = 1000, 100, 10, 1$.} 
\label{fig.full_model}
\end{figure}

Next, we run the full dynamic model for p+p collisions: the medium is initialized with quark-gluon strings of radius $r_\mathrm{str} = 0.25$~fm that evolve in rapidity for $t_\mathrm{max} = 1.5$~fm/\textit{c} and are initially scattered within the transverse plane. This setup generates spatial and temporal fluctuations in the 3D string density, resulting in bin-by-bin variations of $\varepsilon_\mathrm{cell}$, $T_\mathrm{cell}$, and $\mu_\mathrm{cell}$. These varying local conditions strongly alter the momentum loss rate of HF quarks at each propagation step. Furthermore, with HF quarks initialized at various sampled $X-Y$ positions, the combined effects lead to a broad distribution in escape times and consequently even larger fluctuations in the total momentum loss. The impact of these dynamic conditions is shown in Fig.~\ref{fig.full_model} for various values of the anisotropy parameter $\xi$. 
The dependence of the HF quark momentum loss on its initial $p_T$ and on the anisotropy parameter $\xi$ is consistent with the simplified model (Fig.~\ref{fig.simple_loss_variation_of_xi}), although the y-axis scale is a factor of two smaller. The observed reduction in average momentum loss, as discussed, results from the convolution of realistic conditions: the string density along the path is typically much less than the maximum, $k_\mathrm{cell} = 100$, and the time spent in the medium is shorter than the maximum $t_\mathrm{max} = 1.5$~fm/\textit{c}, which differs from the idealized brick tests (see Fig.~\ref{fig.simple_loss_variation_of_time} and~\ref{fig.simple_loss_variation_of_N}). 

Notably, our calculated momentum loss is approximately two orders of magnitude smaller than the value reported by EPOS4HQ~\cite{Zhao:2024oma} for p+p collisions which assumes a hydrodynamic stage in the high-multiplicity p+p events and it dominates the HF energy loss. However, even in the more clear context of nucleus-nucleus collision simulations, the mechanism of energy loss accumulation remains highly contested. For instance, the Glasma approach~\cite{Carrington:2021dvw} predicts pre-equilibrium momentum loss of heavy-flavor quark of the same order of magnitude as the one in the thermal phase. In contrast, a full IP-Glasma + MARTINI + MUSIC + UrQMD simulation for Pb+Pb collisions~\cite{Singh:2025duj} indicates that while early-time dynamics may significantly impact heavy-flavor quark momentum broadening, this effect is largely washed out by the subsequent evolution, leaving no measurable signal in final-state observables.

\section{Summary and outlook}
In this work, we study momentum damping of charm quark propagating through a non-homogeneous environment of gluon-populated color flux tubes. Under the assumptions of an ideal thermal gluon bath and purely longitudinal medium dynamics, we find a significantly smaller elastic momentum loss for the heavy-flavor quark in p+p collisions at $\sqrt{s} = 5.02$ TeV compared to the EPOS4HQ result~\cite{Zhao:2024oma}. However, this is not an apple-to-apple comparison as the latter framework includes both elastic and inelastic collisions in the core-corona picture and omits a pre-equilibrium stage. In turn, we show that its duration, along with the medium density, has a severe impact on heavy probe momentum loss, which additionally fluctuates due to variations in the path length of heavy-flavor quarks and the intermittent structure of the medium. Furthermore, we demonstrate that the heavy-flavor quark's momentum loss decreases with increasing momentum anisotropy of the medium particles, as the effective transverse momentum transferred and gluon density decrease with the anisotropy level, given a fixed energy density. Using the latter is inspired by the successful Romatschke--Strickland parametrization of anisotropic hydrodynamics~\cite{Zhao:2025jwf} and seems to be relevant for capturing the degree of non-equilibrium intrinsic to the color flux-tube medium.

A future version of our model is planned to include radiative energy loss for heavy-flavor quarks, along with hadronization and final-state rescattering. This extension is particularly interesting because EPOS4HQ's description of p+p data~\cite{Zhao:2023ucp} shows room for improvement within current experimental uncertainties.

\section*{Acknowledgements}

DP acknowledges full support of Shui Mu scholarship of Tsinghua University. SS is supported by Tsinghua University under grant Nos. 04200500123, 531205006, 533305009. EA acknowledges Saint-Petersburg State University for a research project 103821868.

\section*{ORCID}

\noindent Daria Prokhorova - \url{https://orcid.org/0000-0003-3726-9196}

\noindent Shuzhe Shi - \url{https://orcid.org/0000-0002-3042-3093}

\noindent Evgeny Andronov - \url{https://orcid.org/0000-0003-0437-9292}

\bibliographystyle{ws-ijmpe}
\bibliography{sample}

\end{document}